\documentclass[12pt]{article}

\newcommand{\m}{\mathrm}
\newcommand{\be}{\begin{equation}}
\newcommand{\ee}{\end{equation}}
\newcommand{\ba}{\begin{eqnarray}}
\newcommand{\ea}{\end{eqnarray}}

\usepackage{graphicx}
\usepackage{amssymb}
\usepackage{amsmath}
\usepackage[T1]{fontenc} 
\usepackage[ansinew]{inputenc} 
\usepackage[nosort]{cite}
\newcommand{\inbar}{\vrule height1.57ex width.4pt depth0pt}
\newcommand{\SW}{\relax{\hbox{$\ \inbar\kern-.285em{\rm S}$}}}

\usepackage{hyperref}
\usepackage{url}
\usepackage{xcolor}
\usepackage{color}
\definecolor{vividviolet}{rgb}{0.62, 0.0, 1.0}
\definecolor{amaranth}{rgb}{0.9, 0.17, 0.31}
\definecolor{palatinateblue}{rgb}{0.15, 0.23, 0.89}
\definecolor{brightpink}{rgb}{1.0, 0.0, 0.5}

\hypersetup{ linktoc=all,
    colorlinks, linkcolor={palatinateblue},
    citecolor={brightpink}, urlcolor={amaranth}
}
\newcommand{\changeurlcolor}[1]{\hypersetup{urlcolor=#1}}
\graphicspath{{Images/}}

\usepackage{caption}
\DeclareCaptionJustification{justified}{\leftskip=0pt \rightskip=0pt \parfillskip=0pt plus 1fil}
\begin{document}
\thispagestyle{empty}
\begin{center}

\null \vskip-1truecm \vskip2truecm

{\Large{\bf \textsf{Event Horizon Wrinklification}}}

{\large{\bf \textsf{}}}

{\large{\bf \textsf{}}}

\vskip1truecm

{\large \textsf{Brett McInnes$^a$ and Yen Chin Ong$^{b,c}$}}

\vskip1truecm

\textsf{\\ a. Department of Mathematics, National University of Singapore,  Singapore 119076\footnote{matmcinn@nus.edu.sg}}

\textsf{b. Center for Gravitation and Cosmology, College of Physical Science and Technology, Yangzhou University, Yangzhou 225009, China\footnote{ycong@yzu.edu.cn}}\\
\textsf{c. School of Aeronautics and Astronautics, Shanghai Jiao Tong University, \\Shanghai 200240, China}\\

\end{center}
\vskip1truecm \centerline{\textsf{ABSTRACT}} \baselineskip=15pt
\medskip
The possible existence of stable black holes with entropies larger than the corresponding Schwarzschild black hole has been discussed extensively. The recently proposed ``rough'' black holes provide a concrete example of this. The fear is that, in accordance with the Second Law of thermodynamics, the familiar smooth-skinned black holes might spontaneously ``wrinklify'' into such an object. We show that this fear is to some extent justified, in the sense that AdS black holes with more entropy than the AdS-Schwarzschild black hole of the same mass do exist.

\newpage

\addtocounter{section}{1}
\section* {\large{\textsf{1. Preventing Wrinkles}}}
The idea that the entropy of a black hole is proportional to its horizon area \cite{kn:wald,kn:wall} has been one of the most influential and stimulating, yet puzzling, insights of recent fundamental physics. The most well-known puzzle is precisely the fact that the entropy is computed from the \emph{area}: this was the first indication that lower-dimensional physics can encode apparently higher-dimensional data.

A less-appreciated, but potentially extremely important, puzzle is the following, recently stated clearly in \cite{kn:supe}. The exact, or indeed even the approximate, round spherical geometry of a Schwarzschild black hole is extremely non-generic among all possible geometries for a black hole with given parameters and with an event horizon with the topology of a two-sphere. (As we will see, the stipulation of ``given parameters'' is far from trivial, but let us leave that to one side for the present.) A generic event horizon geometry would be ``wrinkled'', and would have a far larger area. From a thermodynamic point of view, then, the round spherical geometry apparently corresponds to an extremely \emph{low} entropy state, and of course the identification of entropy with a multiple of the area confirms this intuition. One would then expect that there should exist stable, ``wrinkly'' black hole states of higher entropy, with the same ``given parameters''.

The puzzle is that, in accordance with the Second Law, round event horizon geometries should tend to evolve towards those higher-entropy states \cite{kn:supe}: the familiar Schwarzschild geometry should be unstable  to what we can call ``\emph{wrinklification}''.

One can try to argue against the existence of such stable, yet wrinkled, high-entropy states by invoking ``no-hair'' principles. The wrinkled black hole would be smoothed out by gravitational radiation; the resulting round black hole would have lower entropy, but the entropy of the entire system, including the radiation, is (presumably) larger.

However, it has recently been proposed \cite{kn:bar} that black holes might exist with such extremely wrinkled surfaces that the event horizon is like (for example) a triangular Koch fractal surface \cite{kn:koch}, and it is not clear that such a surface can be smoothed out\footnote{Note that other violations of the no-hair principle have of course been proposed; recently \cite{kn:sunny} it has even been suggested that such violations might ultimately be directly observable.}. If such objects can exist, then, according to the argument above, (potentially unlimited) wrinklification might well be the fate of \emph{all} black holes. This does not seem plausible.

Unlimited wrinklification might be prevented in a variety of ways, but the most direct approach would be to argue that, among \emph{all} stable black holes which can reasonably be compared with each other (in a sense to be clarified), \emph{the Schwarzschild black hole has the largest entropy}. This surprising claim is the physical interpretation of the ``reverse isoperimetric inequality'' \cite{kn:cvetgib}. If this is true, then wrinkled black holes cannot be stable; they will always revert back to the round Schwarzschild geometry, by ``radiating away'' the wrinkles in the manner described above. (The reverse isoperimetric inequality applies to the asymptotically AdS case, and this is the only case we study here. The hope is that at least general lessons we might learn from this case might be relevant to the asymptotically flat situation.)

The discussion in \cite{kn:cvetgib} is based on the idea (see \cite{kn:chemistry} for a review) that the AdS cosmological constant should be treated as a thermodynamic variable, namely the pressure. From this point of view, the ``set of comparable black holes'' mentioned above is a set with a common value of a quantity conjugate to the pressure, called the ``\emph{thermodynamic volume}''. That is, the reverse isoperimetric inequality states that the AdS-Schwarzschild black hole has the largest entropy among all black holes with the same thermodynamic volume.

In \cite{kn:superent} it was claimed that a certain class of asymptotically AdS black holes violate the reverse isoperimetric inequality. These ``superentropic'' black holes are constructed from the AdS-Kerr geometry by taking the limit $\mathcal{A} \rightarrow L$, where $\mathcal{A}$ is the specific angular momentum (angular momentum per unit mass) and $L$ is the background AdS length scale. These are very unusual objects; for example, the event horizon is non-compact but of finite area. Recently their status as genuine counter-examples to the reverse isoperimetric inequality has however been called into question \cite{kn:supe}, so the existence of superentropic black holes (in this sense) is unsettled\footnote{Charged BTZ black holes also violate the reverse isoperimetric inequality \cite{1509.05481}. However in this work we focus on 5-dimensional AdS since one can use applied holography (of quark-gluon plasma) to help us understand some aspects of the black holes \cite{kn:98}.}. For reasons we are about to explain, we will refer to a putative black hole with more entropy than the AdS-Schwarzschild black hole of the same thermodynamic volume as being ``superentropic with respect to thermodynamic volume''.

The Second Law only requires evolution towards states with higher entropy when it is possible for conservation laws to be respected. In the case at hand, it is not clear why thermodynamic volume would have to be conserved in the course of wrinklification. Thus it is not clear that, even if one can rule out the existence of black holes which are superentropic with respect to thermodynamic volume, one will thereby rule out limitless wrinklification.

In the context of wrinklification, then, it seems more natural to fix some conserved quantity, the obvious choice being the \emph{mass}: that is, one could compare black holes with different degrees of wrinklification but the same mass. The attempt to rule out unlimited wrinklification would then take the form of trying to establish that \emph{the AdS-Schwarzschild black hole has a higher entropy than any other stable black hole of the same mass}. This is very natural, since (for example) it is easy to show that an AdS-Schwarzschild black hole has a higher entropy than an AdS-Reissner-Nordstr\"{o}m black hole \emph{of the same mass}. A black hole with more entropy than the AdS-Schwarzschild black hole of the same mass will be described as being ``superentropic with respect to mass''. If we can show that \emph{stable} objects of this kind cannot exist, then this does rule out unlimited wrinklification.

Here we will show that wrinklification \emph{cannot} be ruled out in this way: we will exhibit a concrete example of an AdS$_5$-Kerr black hole (obtained by taking $\mathcal{A} > L$, instead of $\mathcal{A} \rightarrow L$ as above) which is stable yet superentropic with respect to mass. (It is however \emph{not} superentropic with respect to thermodynamic volume; this is also interesting, as it establishes that the two ways of being superentropic are indeed quite different.) We stress that this does not mean that wrinklification \emph{is} possible: it just means that this possibility cannot be dismissed in the most straightforward manner. It must be admitted, however, that with this route blocked, it is certainly not easy to see how wrinklification can be prohibited.

We begin with a brief description of the black holes in question.

\addtocounter{section}{1}
\section* {\large{\textsf{2. Transunital Black Holes}}}
We are interested in a special case of the AdS$_5$-Kerr metric \cite{kn:hawk,kn:cognola,kn:gibperry}, for a singly rotating, uncharged black hole, which takes the form
\begin{flalign}\label{A}
g\left(\m{AdS_5K}\right)\; = \; &- {\Delta_r \over \rho^2}\left[\,\m{d}t \; - \; {a \over \Xi}\,\m{sin}^2\theta \,\m{d}\phi\right]^2\;+\;{\rho^2 \over \Delta_r}\m{d}r^2\;+\;{\rho^2 \over \Delta_{\theta}}\m{d}\theta^2 \\ \notag \,\,\,\,&+\;{\m{sin}^2\theta \,\Delta_{\theta} \over \rho^2}\left[a\,\m{d}t \; - \;{r^2\,+\,a^2 \over \Xi}\,\m{d}\phi\right]^2 \;+\;r^2\cos^2\theta \,\m{d}\psi^2 ,
\end{flalign}
where
\begin{eqnarray}\label{B}
\rho^2& = & r^2\;+\;a^2\cos^2\theta, \nonumber\\
\Delta_r & = & \left(r^2+a^2\right)\left(1 + {r^2\over L^2}\right) - 2M,\nonumber\\
\Delta_{\theta}& = & 1 - {a^2\over L^2} \, \cos^2\theta, \nonumber\\
\Xi & = & 1 - {a^2\over L^2}.
\end{eqnarray}
Here $L$ is the background AdS$_5$ curvature length scale, $t$ and $r$ are as usual, and $\phi$ and $\psi$, and $\theta$ are Hopf coordinates on the topological three-sphere; $M$ and $a$ are geometric parameters describing the size and shape of the event horizon. The physical mass $\mathcal{M}$ is determined by both of these parameters.

It will be useful to use a dimensionless version of $\mathcal{M}$, defined as
\begin{equation}\label{C}
\mu \equiv {8\ell_{\textsf{P}}^3\mathcal{M}\over \pi L^2},
\end{equation}
where $\ell_{\textsf{P}}$ is the bulk Planck length. This dimensionless mass is related to the geometric parameter $M$ by
\begin{equation}\label{D}
\mu \;=\;{2M\over L^2}\,\left({2 + \Xi \over \Xi^2}\right).
\end{equation}
The specific angular momentum $\mathcal{A}$ is given by
\begin{equation}\label{E}
\mathcal{A}\;=\;{2 a \over 3 - \left(a^2/L^2\right)}.
\end{equation}
When Censorship holds, the event horizon is located at $r = r_{\textsf{H}}$, the largest real solution of the equation
\begin{equation}\label{F}
\left(r_{\textsf{H}}^2+a^2\right)\left(1 + {r_{\textsf{H}}^2\over L^2}\right) - 2M\;=\;0.
\end{equation}
We see that a near-extremal black hole of this kind has $r_{\textsf{H}} \approx 0$.

Censorship for these black holes \cite{kn:98} excludes, for each value of $\mu$, a \emph{finite} range of values for $\mathcal{A}/L$, extending below and above unity. The black holes corresponding to values above unity were called ``transunital'' black holes in \cite{kn:100}, and these are the objects of interest to us here. In detail, the condition for Censorship to hold is that $\mathcal{A}/L$ must satisfy either
\begin{equation}\label{G}
\mathcal{A}/L \;<\; \Gamma_{\mu}^- \;<\;1,
\end{equation}
or
\begin{equation}\label{H}
1\;<\; \Gamma_{\mu}^+ \;<\; \mathcal{A}/L,
\end{equation}
where $\Gamma_{\mu}^+$ and $\Gamma_{\mu}^-$ are given by
\begin{equation}\label{I}
\Gamma_{\mu}^{\pm}\;=\;2\,\sqrt{2}\,\sqrt{\mu + 1}\,{\sqrt{3 + 2\mu \pm \sqrt{9 + 8\mu}}\over 3 + 4\mu \mp \sqrt{9 + 8\mu}}.
\end{equation}
Thus, in principle (that is, leaving aside any instabilities), transunital black holes can have arbitrarily large specific angular momenta (for a fixed mass) without violating Censorship.

The entropy of a transunital black hole is given by \cite{kn:100}
\begin{equation}\label{J}
S_{\textsf{T}} \;=\; {\pi^2\over 2 \ell_{\textsf{P}}^3}\,{r^2_{\textsf{H}} + a^2\over |\Xi|}\,r_{\textsf{H}}{L^2\over a^2}.
\end{equation}
Among transunital black holes, the cases nearest to being extremal are those with the \emph{lowest} specific angular momenta. As extremality is approached, $|\Xi|$ becomes smaller but not arbitrarily small\footnote{The exception would be if the mass is simultaneously taken to diverge, which is not relevant here since it does not happen in either of the cases we consider.}, whereas $r_{\textsf{H}}$ does tend to zero in this limit. Therefore, the entropy tends to zero as extremality is approached, that is, as the temperature drops towards zero. This is in accord with the Nernst-Simon formulation of the Third Law of thermodynamics (which, in fact, most black holes fail to satisfy).

A near-extremal transunital AdS$_5$-Kerr black hole, therefore, certainly has a smaller entropy than the corresponding AdS$_5$-Schwarzschild black hole, whether one interprets ``corresponding'' to mean ``having the same thermodynamic volume'' or ``having the same mass''. On the other hand, as we move away from extremality, by \emph{increasing} the specific angular momentum, one expects the entropy to increase indefinitely; we will actually prove this in the case where the mass is fixed. Thus, it is possible that a transunital black hole can become superentropic, \emph{if} $\mathcal{A}$ can increase sufficiently without causing the black hole to become unstable.

We now study this possibility in detail, for the two definitions of being superentropic.

\addtocounter{section}{1}
\section* {\large{\textsf{3. Transunital Black Holes Can Be Superentropic With Respect to Mass}}}
In this Section, we wish to explore the possible existence of AdS$_5$-Kerr black holes with more entropy than the AdS$_5$-Schwarzschild black hole of the same \emph{mass}.

As we have seen, from the point of view of wrinklification, the key property of transunital black holes is that their entropy increases with their specific angular momentum. When the mass is fixed, the underlying physics of this phenomenon can be understood by computing the angular velocity (relative to the coordinates we are using) of a massless zero-angular-momentum particle fixed on the horizon (the ``angular velocity of the horizon''):
\begin{equation}\label{K}
\Omega\;=\;{a\,\Xi\over r_{\textsf{H}}^2 + a^2}.
\end{equation}
Since $\Xi$ is negative in the transunital case, this has the opposite sign to that of the angular momentum. If, therefore, one increases the angular momentum of the black hole without changing its mass, then, by the First Law of (black hole) thermodynamics, the entropy must indeed increase.

We need to compute how by how much the specific angular momentum has to increase in order for the entropy to exceed the AdS$_5$-Schwarzschild value.

Now equation (\ref{F}) allows us to compute $r_{\textsf{H}}$ given $a$ and $M$. Equations (\ref{D}) and (\ref{E}) then allow us to express $r_{\textsf{H}}$ in terms of $\mu$ and $\mathcal{A}$. Therefore, equation (\ref{J}) permits us to express $S_{\textsf{T}}$ in terms of $\mu$ and $\mathcal{A}$.

The entropy of an AdS$_5$-Schwarzschild black hole\footnote{Note carefully that, from equation (\ref{D}), the dimensionless mass in the AdS$_5$-Schwarzschild case is not $M/L^2$ but rather $6M/L^2$.} as a function of (this same) $\mu$ is
\begin{equation}\label{L}
S_{\textsf{S}} \;=\; {\pi^2\over 2 \ell_{\textsf{P}}^3}\,\left[{1\over 2}\left(-1 + \sqrt{1 + {4\mu \over 3}}\right)\right]^{3/2}.
\end{equation}
Finally, then, the ratio $S_{\textsf{T}}/S_{\textsf{S}}$ can be computed and expressed explicitly as a function of $\mu$ and $\mathcal{A}$.

\begin{figure}[!h]
\centering
\includegraphics[width=0.7\textwidth]{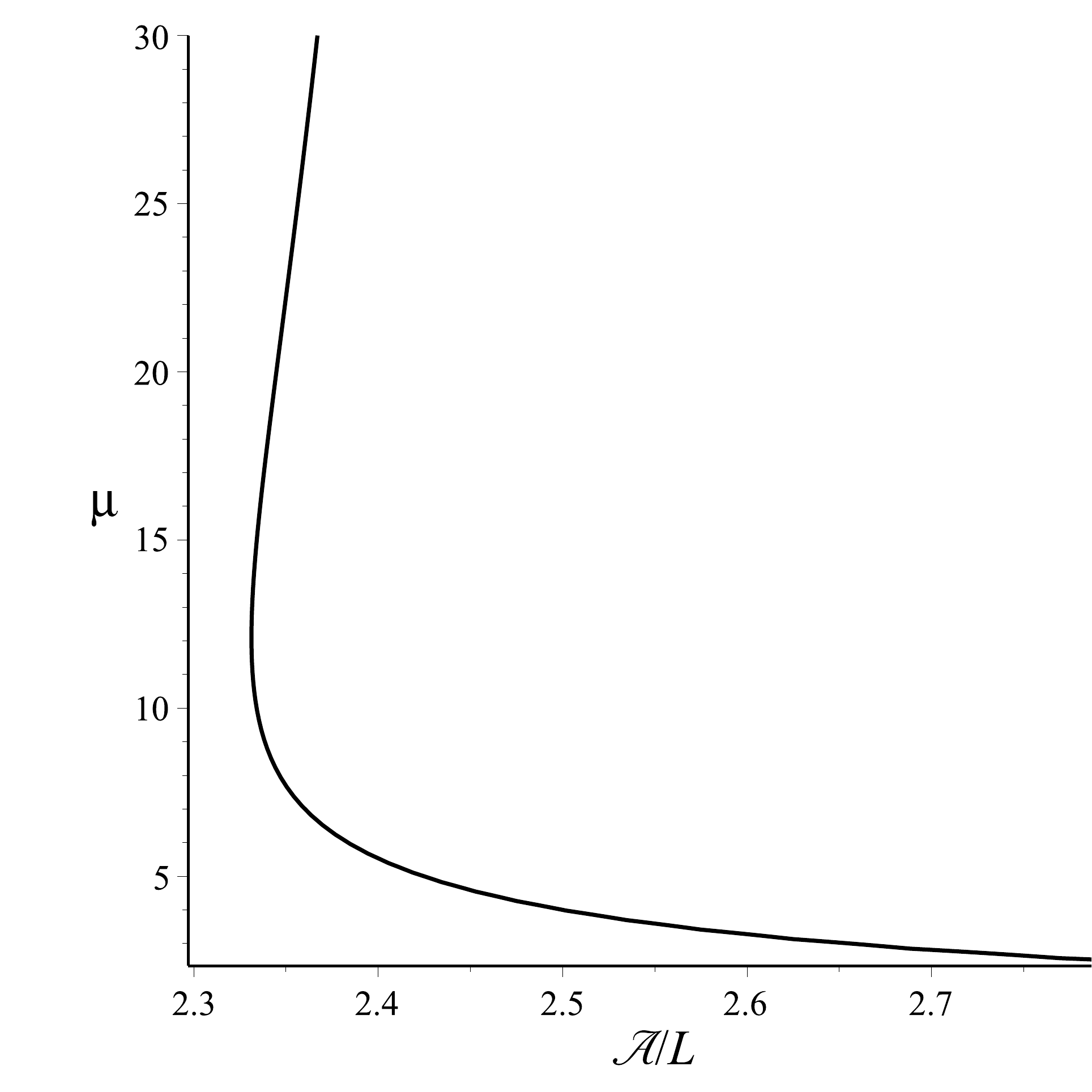}
\caption{The locus $S_{\textsf{T}}/S_{\textsf{S}} = 1$, the boundary between parameter values for transunital black holes which are superentropic with respect to mass and those which are not. \label{stss}}
\end{figure}
While it is possible to write down this function explicitly, it is too complicated to yield any insight. Instead, we set $S_{\textsf{T}}/S_{\textsf{S}} = 1$ and graph the resulting function of a single variable. This is shown as Figure (\ref{stss}).

All points in this diagram lying to the right of the curve correspond to pairs $\left(\mathcal{A}/L, \mu\right)$ with $S_{\textsf{T}}/S_{\textsf{S}} > 1$: these are the superentropic black holes we seek. Notice that superentropic black holes of this kind cannot exist, for any value of the mass, unless $\mathcal{A}/L$ is at least as large as about $2.33$.

For clarity, let us focus on a specific case. We choose
\begin{equation}\label{L}
\left(\mathcal{A}/L, \mu\right) = \left(2.692, \, 4.514\right);
\end{equation}
other (nearby) choices are possible, but do not lead to different conclusions. We have
\begin{equation}\label{M}
S_{\textsf{T}}/S_{\textsf{S}}\left(2.692, \, 4.514\right) \;\approx\; 1.313;
\end{equation}
so this object is clearly superentropic with respect to mass, when the dimensionless mass is fixed at $4.514$.

Let us verify that Censorship is satisfied here. We have, from equations (\ref{I}),
\begin{equation}\label{N}
\Gamma_{\mu}^{+}\left(\mu = 4.514\right)\;\approx\; 2.005,
\end{equation}
and clearly the inequality (\ref{H}) is satisfied here with $\mathcal{A}/L = 2.692$.

Thus, the parameter values given in (\ref{L}) do give us a well-defined black hole, superentropic with respect to mass. The question now is whether this object is stable.

\addtocounter{section}{1}
\section* {\large{\textsf{4. Questions of Stability}}}
The black hole we have just described has a very high specific angular momentum, and that might easily cause it to be unstable in one manner or another. Let us consider whether that is so.

\subsubsection*{{\textsf{4.1 Stability Against Superradiance }}}
Rapidly rotating black holes are often unstable due to the loss of energy and angular momentum to superradiant modes \cite{kn:super}. In the AdS context, this will happen unless \cite{kn:hawkreall}
\begin{equation}\label{O}
{{a\over L}\,\left(1 + {r_{\textsf{H}}^2\over L^2}\right)\over {r_{\textsf{H}}^2\over L^2} + {a^2\over L^2}}\;<\;1.
\end{equation}
As above, we can regard the left side of this inequality as a function of $\mathcal{A}/L$ and $\mu$; regarding it as such, we call it $\Sigma\left(\mathcal{A}/L, \mu\right)$. We find that
\begin{equation}\label{P}
\Sigma\left(2.692, \, 4.514\right) \;\approx\; 0.870.
\end{equation}
Thus, (\ref{O}) is satisfied, and the superentropic black hole with parameter values given in (\ref{L}) is in no danger of a superradiant instability.

\subsubsection*{{\textsf{4.2 Stability Against Fragmentation }}}
Rapidly rotating black holes are also in danger of \emph{fragmentation} \cite{kn:empmy,kn:pau}. Again, the criterion for this is thermodynamic: one asks whether a pair of black holes, each with maximal entropy $S_1$ given its mass, and with the same total energy as a given black hole, has a total entropy greater than the entropy, $S_0$, of the original system. If so, then one expects the system to evolve accordingly, that is, to split into two or more fragments.

In \cite{kn:100} the ratio of the two entropies was computed for transunital black holes. Evaluating it at the parameter values given in (\ref{L}), we have
\begin{equation}\label{Q}
{2S_1\over S_0}\left(2.692, \, 4.514\right) \;\approx\; 0.585.
\end{equation}
Even allowing for the many approximations in this computation, this is far below unity, and so we can conclude that our black hole is stable against fragmentation.

\subsubsection*{{\textsf{4.3 Stability Against the Seiberg-Witten Effect }}}
When asymptotically AdS black holes are embedded in string theory, one finds \cite{kn:96} that the angular momentum of the black hole tends to destabilize it, by triggering pair-production of branes \cite{kn:seiberg,0409242}. This happens when $\mathcal{A}$ is sufficiently large relative to $L$.

In \cite{kn:96} it was shown that this Seiberg-Witten effect can be avoided if
\begin{equation}\label{R}
\mathcal{A}/L\;\leq \;2\,\sqrt{2}\;\approx \; 2.828.
\end{equation}
This condition is of course satisfied by our black hole, with $\mathcal{A}/L = 2.692$. \newline

\begin{figure}[!h]
\centering
\includegraphics[width=0.8\textwidth]{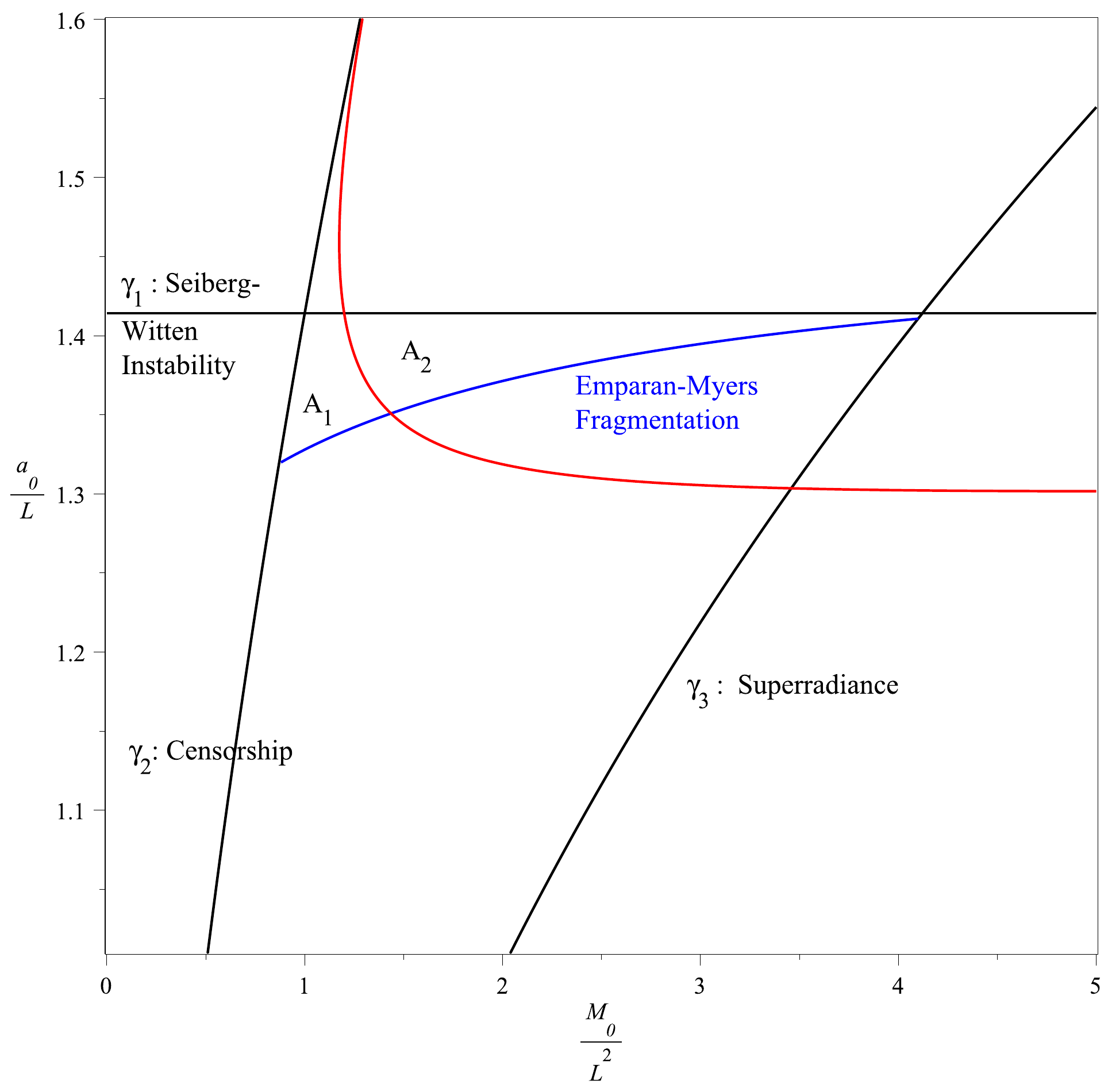}
\caption{Constraints on transunital black holes. Only the regions labelled $A_1$ and $A_2$ correspond to stable black holes (see text for detail). Region $A_2$ corresponds to stable black holes that are superentropic with respect to mass. \label{region}}
\end{figure}

In Fig.(\ref{region}), we show the range of parameters in which the black holes are stable. The curves $\gamma_1, \gamma_2, \gamma_3$ bound a region in which black holes exist (Censorship is satisfied) and are stable against superradiance as well as the Seiberg-Witten effect. Specifically, the region to the left of $\gamma_2$ corresponds to objects lacking a horizon (naked singularities), that above $\gamma_1$ to black holes unstable against brane pair production, and that to the right of $\gamma_3$ to black holes unstable due to superradiance. Below the blue curve, one has black holes unstable against fragmentation (the subscript ``0'' for $a_0$ and $M_0$ indicates, in this context, the initial configuration of the black hole -- if thermodynamically unstable it will transit into another configuration. See \cite{kn:100} for detail).

Thus, only the regions denoted by $A_1$ and $A_2$  are physically relevant. The region to the right of the red curve, $A_2$, corresponds to stable black holes which are superentropic with respect to mass.

\addtocounter{section}{1}
\section* {\large{\textsf{5. Transunital Black Holes are Never Superentropic With Respect to Thermodynamic Volume}}}
In this Section, we wish to explore the possible existence of AdS$_5$-Kerr black holes with more entropy than the AdS$_5$-Schwarzschild black hole of the same \emph{thermodynamic volume}.

We now show that although AdS$_5$-Kerr transunital black holes can be superentropic with respect to mass, they are nevertheless never superentropic with respect to \emph{thermodynamic volume}, so these two notions are indeed distinct.

We mentioned earlier that there are two kinds of AdS$_5$-Kerr black holes: the transunital ones we have been discussing, and those --$\,$ we can call them \emph{cisunital} --$\,$ satisfying the inequality (\ref{G}). It turns out that the expression for the entropy, $S_\textsf{T}$, differs from the cisunital one by a factor of $L^2/a^2$. However, since $\mathcal{M}$ has the same expression for both transunital and cisunital cases,
it follows from the First Law that the product $ST$ is identical for both cases.
The thermodynamic volume thus also has an identical form, and is given by \cite{kn:chemistry}
\begin{equation}
V(M,a,L)=\frac{r_\textsf{H} A}{4} + \frac{\pi^2 a^2 M}{3 \Xi^2},
\end{equation}
where $A$ denotes the horizon area.

It suffices to check that the reverse isoperimetric inequality \cite{kn:superent},
\begin{equation}\label{rii}
R(M,a,L):=\left[\frac{4 V(M,a,L)}{\omega_3}\right]^{\frac{1}{4}}\left[\frac{\omega_3}{A(m,a,L)}\right]^{\frac{1}{3}} < 1,
\end{equation}
holds in the region of interest (the ranges of parameters such that the black holes are stable). 
Here $\omega_3$ is the area of the unit 3-sphere.
This is indeed so as shown in Fig.(\ref{region}). We remark that all cisunital AdS$_5$-Kerr black holes satisfy the reverse isoperimetric inequality \cite{kn:superent}.

\begin{figure}[!h]
\centering
\includegraphics[width=1.1\textwidth]{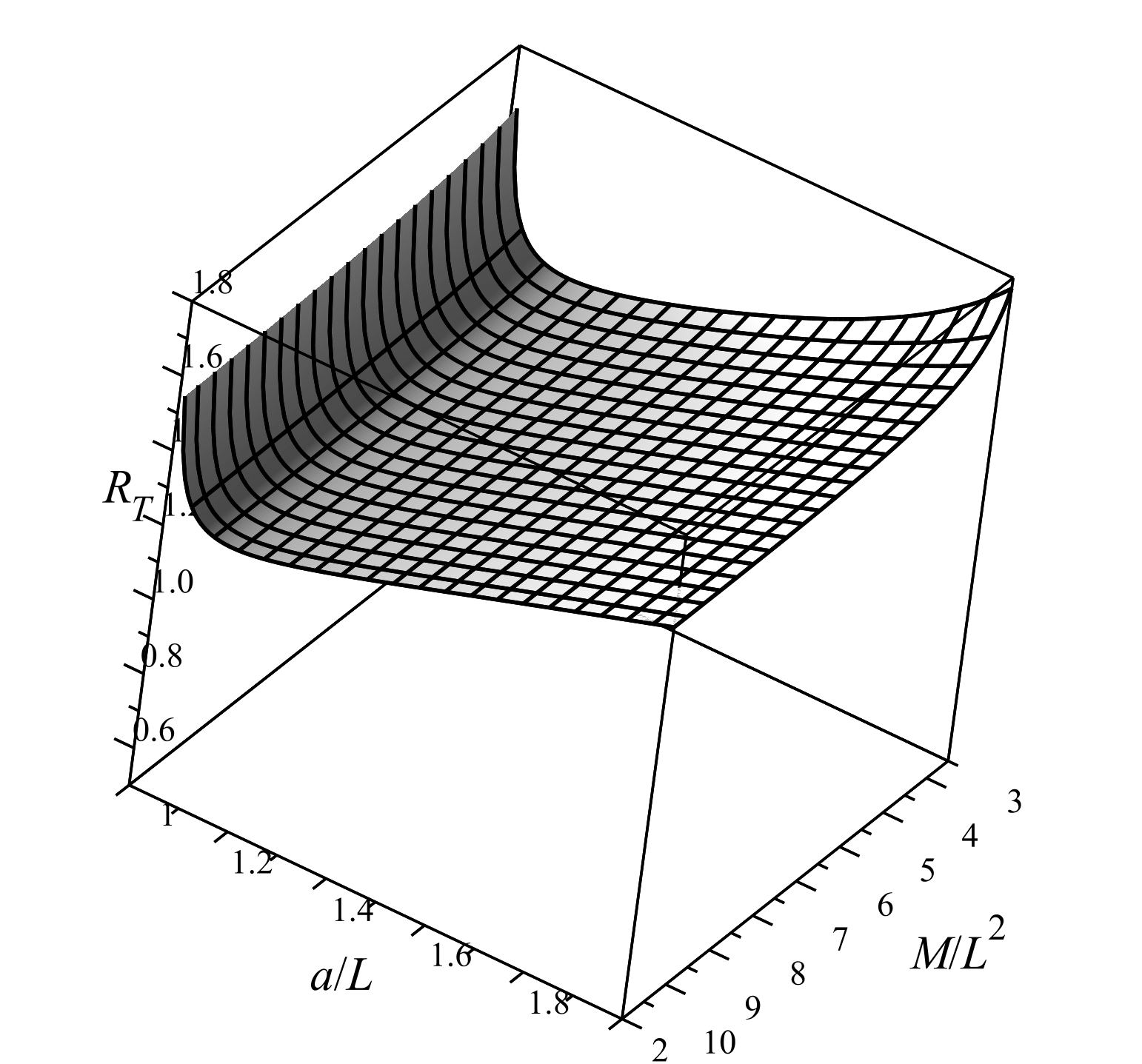}
\caption{The quantity $R$ (as defined in Eq.(\ref{rii})) is larger than 1, indicating that the black holes are not superentropic with respect to the thermodynamic volume. The subscript ``T'' emphasizes that the black hole is transunital. \label{region}}
\end{figure}

\addtocounter{section}{1}
\section* {\large{\textsf{6. Conclusion: Wrinkles May Be Unavoidable}}}
In summary: we have constructed a black hole, described by the physical parameters given in (\ref{L}), which satisfies Censorship, which is surprisingly stable in a multitude of ways (ranging from classical thermodynamics to semi-classical effect and stringy brane emission) despite its high specific angular momentum, which is not superentropic with respect to the thermodynamic volume, but which \emph{is} superentropic with respect to mass. We argue that, in the specific context of wrinklification, it is this latter definition that is the relevant one.

We see, then, that there probably cannot be any universal ban on objects with larger entropies than the AdS$_5$-Schwarzschild black hole. This does not, of course, prove that there exist stable, extremely wrinkled black holes of the kind discussed in \cite{kn:bar}, into which the familiar smooth-skinned black holes must inevitably age. It does however indicate that this possibility should be considered very carefully.

\addtocounter{section}{1}
\section*{\large{\textsf{Acknowledgement}}}
BMc is grateful to Dr. Soon Wanmei for useful discussions. YCO thanks the National Natural Science Foundation of China (No.11705162, No.11922508) and the Natural Science Foundation of Jiangsu Province (No.BK20170479) for funding support.

\end{document}